\documentclass[prd,10pt,twoside,english,nofootinbib,preprintnumbers,aps,onecolumn,notitlepage]{revtex4-2}
\usepackage{euscript}
\usepackage{amsthm}
\usepackage{graphicx}
\usepackage{amsfonts}
\usepackage{amsmath}
\usepackage{amssymb}
\usepackage{mathtools}
\usepackage[normalem]{ulem}
\usepackage{fancyhdr}
\usepackage{epsfig}
\usepackage{color}
\usepackage{esint}
\usepackage{soul}
\usepackage{calc}
\usepackage{enumerate}
\usepackage{enumitem}
\usepackage{afterpage}
\usepackage{physics}
\usepackage[papersize={8.5in,11in}]{geometry}
\newcommand\redsout{\bgroup\markoverwith{\textcolor{red}{\rule[0.5ex]{2pt}{0.4pt}}}\ULon}

\usepackage{rotating}
\usepackage{array}
\usepackage{booktabs}
\usepackage{verbatim}
\usepackage[dvipsnames]{xcolor}
\usepackage{multirow}

\usepackage[colorlinks=true,
citecolor=red,
linkcolor=blue,
urlcolor=violet,
filecolor=cyan,
backref=false]{hyperref}

\usepackage{svg}

\allowdisplaybreaks

\newcommand{\be}{\begin{equation}}
\newcommand{\ee}{\end{equation}}

\begin{document}
\title{Extremal isolated horizons of the NUT type\footnote{This work is dedicated to the memory of Jerzy Lewandowski, whose guidance as a teacher and mentor left a lasting impact not only on our research, but also on us as individuals.}}
\author{Eryk Buk}\email{Eryk.Buk@fuw.edu.pl}
\affiliation{Faculty of Physics, University of Warsaw, ul. Pasteura 5, 02-093 Warsaw, Poland}
\author{Denis Dobkowski-Ry{\l}ko}
\email{Denis.Dobkowski-Rylko@ug.edu.pl}
 \affiliation{Institute of Theoretical Physics and Astrophysics, Faculty of Mathematics, Physics and Informatics, University of Gdansk, Wita Stwosza 57, 80-308, Gdańsk, Poland}
 \affiliation{Faculty of Physics, University of Warsaw, ul. Pasteura 5, 02-093 Warsaw, Poland}

\author{Jerzy Lewandowski}
	\affiliation{Faculty of Physics, University of Warsaw, ul. Pasteura 5, 02-093 Warsaw, Poland}
\author{Maciej Ossowski}
	\email{Maciej.Ossowski@uj.edu.pl}
	\affiliation{Faculty of Physics, University of Warsaw, ul. Pasteura 5, 02-093 Warsaw, Poland}
    \affiliation{Institute of Theoretical Physics, Jagiellonian University, Łojasiewicza 11, 30-348 Kraków, Poland}
%



\begin{abstract}
We provide a construction of a new class of axisymmetric extremal isolated horizons admitting a structure of U(1)-principal fiber bundle over a two-sphere. 
In contrast to the previous examples, the null generators are assumed to be transversal to the bundle fibers. 
We impose the Einstein equations at the horizon and explicitly derive all intrinsic geometries of the extremal horizon, consisting of a two-sphere metric and a rotation 1-form, in the above class. 
The 2-geometries turn out to be equivalent to the classification of conically singular horizons with product topology.
Both the rotating and non-rotating horizons are then embedded in the Plebański-Demiański spacetimes, which naturally admit horizons of this type.
Furthermore, we compare our results with previously obtained solutions to the Einstein vacuum extremal horizon equation with cosmological constant and the solution of Petrov type D equation with transversal bundle structure.
\end{abstract}

\maketitle

\tableofcontents

\section{Introduction}
\subsection{Motivation}
The geometry of Killing's horizons can be reformulated without reference to the spacetimes that surround them.  
It consists of a manifold, a degenerate metric tensor, and a compatible connection.  
Such a space is called an \textit{isolated horizon} \cite{ABF, geometryhorizonsAshtekar_2002, MechanicsIHPhysRevD.64.044016, Ashtekar2004}. 
Its geometry is constrained by equations resulting from the spacetime Einstein's equations \cite{geometryhorizonsAshtekar_2002}. 
Remarkably, in the extremal $\kappa=0$ case, the equations may be stated and solved purely intrinsically, without a reference to the bulk spacetime.
It turns out that many properties of black holes follow from the laws governing the geometry of isolated horizons. 
The isolated horizons program is devoted to the study of these laws \cite{Lewandowski_2003, higherDim, PawLewJez2003, typeD}.  
The symmetry of a Killing horizon carries over to the local symmetry of an isolated horizon, whereas many properties of Killing horizons -- such as extremality -- find their counterparts in the isolated horizon framework.
An example is a beautiful result concerned with the uniqueness of the topology and rotation of the extremal isolated horizon, arising from the assumption of the existence of a global section and the validity of Einstein's vacuum equations in four-dimensional spacetime. 
Indeed, the section must be a topological sphere, the horizon is characterized by intrinsic rigidity, and the rotation vector field is unique \cite{Lewandowski_2003}. 
Those results admit suitable generalization to higher dimensions and to the case of Einstein's equations with non-zero cosmological constant or even with matter fields \cite{higherDim, higherReall, lucietti2009, Lucietti2013}. 
There are still many questions that have not been answered, and many ideas that are waiting to be implemented. 

In the current work, we study topologically non-trivial extremal isolated horizons, i.e., horizons that do not admit a global section.
They have the Hopf bundle topology, or more broadly, a topology of $L(n,1)$ lens space, with a structure of a $U(1)$-principal fiber bundle over a two-sphere. 
The structure group of the bundle gives rise to a cyclic symmetry of the horizon geometry; however, the null curves are transversal to the bundle fibers
\footnote{A different case where null direction is tangent to the fibers was studied in \cite{hopf, LO2}.}. 
These types of horizons occur naturally in the Kerr-NUT-(a)dS spacetimes, and their accelerated generalization with generic values of parameters contained in the Plebański-Demiański family of spacetimes, if the Misner interpretation is assumed as in \cite{LO3PhysRevD.104.024022}. 
The horizons of this type were already considered in the context of the Petrov type D equation \cite{DLO} - another equation constraining the intrinsic isolated horizon geometry. 

Below, we extend this construction to include a previously overlooked case of an extremal isolated horizon with a non-trivial principal bundle structure satisfying the Einstein equations.
Two ingredients of the horizon structure that need to be constrained by the field equations are the axially symmetric metric tensor $g$ on a 2-sphere, with a necessary conical singularity, and a smooth 1-form $\omega$, both defined on a 2-sphere.
We provide all such pairs $(g,\omega)$ satisfying the so-called \textit{Einstein vacuum extremal horizon (EEH) equation}, i.e. a restriction of the $\Lambda$-vacuum Einstein equations to the horizon. 
In the second part of this paper, we compare the obtained family of solutions with previous results.
This includes solutions to the Petrov type D equation obtained by Dobkowski-Ry{\l}ko, Lewandowski, and Ossowski \cite{DLO}, the generic axially symmetric solutions to the EEH found by Lucietti and Kunduri \cite{Lucietti2013}, and the extremal horizon geometry classification by Podolsky and Matejov \cite{PodolskyMatejov2021,PodolskyMatejov2022}.
As all of the above approaches admit conical singularities, their results overlap with our EEH solutions to a degree.
The solutions to the type D equation necessarily contain solutions to the EEH, however, they are not easily distinguishable from the non-extremal ones.
The Podolsky-Matejov solutions were obtained by reducing the Einstein-Maxwell equations to the extremal horizon using the Newman-Penrose formalism. In particular, in the absence of an electromagnetic field, these equations reduce to the EEH equation considered and solved in this work. 
Our solutions for the horizon 2-geometries coincide with theirs for the vanishing electromagnetic field.

However, we wish to stress that our goal is not the classification of spherical, axially symmetric solutions to EEH, but rather providing a detailed construction of a smooth extremal isolated horizon of the $U(1)$-principal fiber bundle over the two-sphere and all intrinsic geometries $(g,\omega)$ admissible by this construction. 
Therefore, we solve the constraints in the manner adapted to this task and explicitly exhibit all horizon geometries $g$ as well as their rotation 1-form potential $\omega$.
Additionally, we analyze the admissible parameter ranges of the resulting solutions and show that the intrinsic geometries $(g,\omega)$ have a proper limit to the smooth solutions obtained by Lewandowski and Buk \cite{Buk}.

\subsection{Vacuum extremal isolated horizons}
An \textit{isolated horizon} is an $(n-1)$-dimensional  manifold $H$ endowed with 
\begin{itemize}
    \item a symmetric tensor ${}^{(n-1)}g$ of rank $(n-2)$
    \item a torsion free covariant derivative ${}^{(n-1)}\nabla$ such that
    \begin{equation}
     {}^{(n-1)}\nabla  {}^{(n-1)}g = 0
      \end{equation}
     \item a nowhere vanishing, null vector field $\ell$ defined up to 
     \begin{equation}
         \ell\mapsto c_0 \ell
         \qquad
         c_0\in\mathbb{R}_{+}
         ,
     \end{equation}
     such that 
     \begin{equation}\label{g1}
        \ell\lrcorner {}^{(n-1)}g=0,
        \quad
        \mathcal{L}_\ell  {}^{(n-1)}g=0,
        \quad
        [\mathcal{L}_\ell,{}^{(n-1)}\nabla]=0,
        \quad\text{and}\quad
        {}^{(n-1)}\nabla_\ell \ell = 0. 
    \end{equation}
\end{itemize}
The rotation $1$-form potential $ {}^{(n-1)}\omega$ is a part of the covariant connection defined on $H$ by the {equality}
\begin{equation}\label{nablaell}
 {}^{(n-1)}\nabla \ell =  {}^{(n-1)}\omega \otimes \ell.
\end{equation} 
It follows that 
\begin{equation}\label{omega1}
    \ell\lrcorner {}^{(n-1)}\omega = 0,
    \quad\text{and}\quad
    \mathcal{L}_\ell{}^{(n-1)}\omega=0.   
\end{equation}

Consider now any local $(n-2)$-dimensional section $S\subset H$, transversal to $\ell$. 
There is a metric tensor $g$ induced on $S$, {which determines metric, torsion free covariant derivative $\nabla$, together with the corresponding Riemann tensor  $R$, and the 1-form $\omega$ -- the pullback of ${}^{(n-1)}\omega$ to $S$}. If the extremal isolated horizon $(H,{}^{(n-1)}g,{}^{(n-1)}\nabla,\ell)$ were embedded in a $n$-dimensional spacetime {${}^{(n)}M$} endowed with metric tensor ${}^{(n)}g$, then the following geometric identity would be true \cite{Hajicek, higherDim, geometryhorizonsAshtekar_2002, Moncrief}
\begin{equation}
{}^{(n)}R_{AB} =  R_{AB} - 2\nabla_{(A}\omega_{B)} - 2\omega_A\omega_B,     
\end{equation}
where ${}^{(n)}R_{AB}$ is the pull back to $S$ of the spacetime Ricci tensor while $R_{AB}$ is the Ricci tensor of the intrinsic metric $g$ of $S$. Therefore, if we assume that the vacuum Einstein equations with cosmological constant hold on $H$, namely 
\begin{equation}
    {}^{(n)}R_{\alpha\beta} \overset{H}{=} \frac{2\Lambda}{n-2}  {}^{(n)}g_{\alpha\beta},
\end{equation}
 then we obtain an equation for the objects defined on $S$ \begin{equation}\label{extremity}
  \nabla_{(A}\omega_{B)} + \omega_A\omega_B -\frac{1}{2}R_{AB} +\frac{\Lambda}{n-2}g_{AB} = 0.    
 \end{equation}
 This equation is satisfied by every local section $S$ of an extremal isolated horizon $H$ embedded in vacuum spacetime {and is referred to as the {\it \textit{Einstein vacuum extremal horizon} (EEH) equation with cosmological constant $\Lambda$.}}
 Since it involves only a structure defined on $H$, it can also be introduced in the case of an unembedded isolated horizon $H$ and provides a definition of an {\it unembedded vacuum extremal isolated horizon}.
 
 Given an extremal isolated horizon $H$ endowed with a degenerate metric tensor ${}^{(n-1)}g$ and a rotation $1$-form $ {}^{(n-1)}\omega$, if the set ${\cal S}$ of the null geodesics, that is the integral curves of the vector field $\ell$, has the Hausdorff topology and a structure of $(n-2)$-dimensional manifold such that the map
 \begin{equation}\label{pi}
   \pi: H\rightarrow {\cal S}   
 \end{equation}
is differentiable, then there is a metric tensor $g$ and a $1$-form $\omega$ differentiable and globally defined on ${\cal S}$, such that 
\begin{equation}\label{pullback}
     {}^{(n-1)}g = \pi^* g
     \quad\text{and}\quad
     {}^{(n-1)}\omega=\pi^* \omega.
\end{equation}

Locally, each section $S$ of the horizon H is diffeomorphic to the base space $\mathcal S$; hence, we do not introduce separate notation for objects defined on either.
Consequently, we attribute the EEH equation (\ref{extremity}) to $(\mathcal{S}, g,\omega)$ \cite{higherDim}.
This case was studied in the literature, and we outline now the main results of the extremal horizon theory.
If ${\cal S}$ is topologically {a} $2$-sphere ({corresponding} to the  spacetime dimension $n=4$), then the equation (\ref{extremity})  has a unique solution for each $\Lambda \not= 0$ \cite{Buk}, and a unique solution up to $g\mapsto b_0 g$ where $b_0\in\mathbb{R}_{+}$, when $\Lambda=0$ \cite{Lewandowski_2003}.
These solutions correspond to the extremal Kerr-(anti-)de Sitter \cite{Buk} and the extremal Kerr black hole {horizon} \cite{Lewandowski_2003}, respectively. A detailed classification of the results in four dimensions ($n=4$), together with their generalization to higher dimensions ($n>4$), is given in \cite{lucietti2009, Lucietti2013}.
If dim\,${\cal S}=2$, however the Euler characteristic of the space of the null generators is negative, i.e. $\chi_E({\cal S})\le 0$, then the only solutions are $\omega=0$ and $g$ of constant curvature, with Ricci scalar $R=2\Lambda$ \cite{DOBK} (see also \cite{KamLew2024, CollDun2025} for similar results applying to generalized extremal horizons equation including quasi-Einstein spaces and contributions from a Maxwell field).  
For every smooth and compact ${\cal S}$ of arbitrary dimension $n-2$,  an intrinsic rigidity theorem was proven, first for non-negative  \cite{Dunajski2023} and later generalized for arbitrary cosmological constant \cite{CDKL}:  if $(g,\omega)$ is a solution to the EEH equation~(\ref{extremity}) on ${\cal S}$, then there exists on ${\cal S}$ a globally defined, non-zero vector field $\xi$, such that
\begin{equation}
    \mathcal{L}_\xi g = 0,
    \quad\text{and}\quad
    \mathcal{L}_\xi \omega = 0,
\end{equation}
provided $\omega$ is not exact. 
This result has been generalized to the case with an electromagnetic field in \cite{CKL} for $n = 4$. Moreover, an alternative proof of the intrinsic rigidity theorem of \cite{Dunajski2023} has been shown in \cite{KamLew2024} and applies for any value of the cosmological constant. 
If $\omega$ is closed, on the other hand, then either $\omega=0$ and $g$ makes ${\cal S}$ an Einstein space (in particular a compact Ricci flat space for vanishing cosmological constant $\Lambda$, constant positive and negative curvature for $\Lambda>0$ and  $\Lambda<0$, respectively) \cite{Chrusciel_2006}, or $\omega$ is nontrivial, which can occur only for $n>4$ (see \cite{KamLew2024, Bahuaud2022} for recent results in higher dimensions).

\bigskip

\noindent{\bf Remark}  If an extremal isolated horizon  $H$ -- or, if $H$ is not geodesically complete, its analytic extension as an abstract three-dimensional manifold -- has the structure of a principal fiber bundle over the space  ${\cal S}$ of the null geodesics, with the structure group defined by the flow of the null vector field $\ell$, a possible non-triviality of the bundle is not an obstacle for $(g,\omega)$ defined above to exist on ${\cal S}$ and be everywhere regular. That is what happens due to the first two equations in (\ref{g1}) and the properties (\ref{omega1}) (for details see \cite{hopf}).

\section{Extremal NUT isolated horizons}
\label{sec:ExtremalNUTIsolatedHorizons}Kerr-NUT spacetimes (possibly with a cosmological constant and charges) can be given a structure of a U(1) principal fiber bundle over $S^2\times \mathbb{R}$. 
A Killing horizon contained in this kind of spacetime inherits the bundle structure. 
However, if the Kerr parameter is not zero, then, generically,  the horizon null generators are not the fibers of the bundle (see Eqs. (84) and (85) in \cite{DLO} for the conditions ensuring simultaneous removal of the conical singularities). 
That is because each horizon has an angular velocity with respect to the infinity.
In particular, this observation applies to the extremal horizons in the Kerr-NUT-(anti-) de~Sitter spacetimes. 
In this case, even though the extremal horizon is smooth as a $3$-dimensional surface, its space of null generators is not a differentiable manifold.
Motivated by this observation, in this section, we construct a family of extremal isolated horizons that are non-trivial in the above sense, and next, we solve the EEH equation for them.    

Consider a (topological) sphere $S^2$ and an axisymmetric metric tensor $g$ defined thereon except for the poles, accompanied by an axisymmetric $1$-form $\omega$ defined and regular on the entire $S^2$. 
We are assuming that $g$ has conical singularities in the poles. 
More specifically, in some spherical coordinates $(\theta,\phi)\in (0,\pi)\times [0,2\pi)$ we have
\begin{equation}
\label{g}
g = g_{\theta\theta}(\theta)d\theta^2 +  g_{\phi\phi}(\theta)d\phi^2 
\quad\text{for}\quad
\theta\notin\{0,\pi\},
\end{equation}
and there are positive numbers $\beta_-$ and  $\beta_+$, such that 
\begin{equation}
    g_- := g_{\theta\theta}(\theta)d\theta^2 +  \beta_-^2g_{\phi\phi}(\theta)d\phi^2,  
\end{equation}
is extendable in a $C^{k+1}$ way ($k\in \mathbb{N}^*$) to the pole  $\theta=0$, while the same is true at the other pole $\theta=\pi$, for 
\begin{equation}
    g_+ := g_{\theta\theta}(\theta)d\theta^2 +  \beta_+^2g_{\phi\phi}(\theta)d\phi^2. 
\end{equation}
Let our sphere be also equipped with an everywhere $C^k$ differential  $1$-form
\footnote{The assumption that $\omega$ is regular at both poles instead of just one, as is the case for $g_+$ and $g_-$ is easily justified: for the extremal horizon, the regularity at one pole implies regularity at the other \cite{DLO}.}
\begin{equation}\label{omega}
    \omega = \omega_\theta(\theta) d\theta + \omega_\phi(\theta) d\phi.
\end{equation} 
Notice that if we use the decomposition
\begin{equation}\label{omegadec}
    \omega = df(\theta) + * dh(\theta),
    \quad
    f,h\in C^{k+1}(S^2), 
\end{equation}
where the Hodge dual is defined by the auxiliary round metric $d\theta^2 + \sin^2(\theta)d\phi^2$, then the first/second term in (\ref{omega}) corresponds to the first/second term in (\ref{omegadec}). The conclusion is that each of the terms on its own is $C^k$ on the entire sphere. 

Based on this data, we will construct an everywhere regular, extremal isolated horizon of $S^3$ topology below. In the next section, we will superimpose the EEH equation on it. 
Consider two $3$-dimensional manifolds:
\begin{itemize}
\item $S^2\setminus \{p_+\}\times S^1$, where $p_+$ is the pole $\theta=\pi$, and 
\item  $S^2\setminus \{p_-\}\times S^1$, where $p_-$ is the pole $\theta=0$,
\end{itemize}
and introduce on $S^1$ a cyclic coordinate $v\in [0,2\pi)$. 
On $S^2\setminus \{p_+\}\times S^1$ consider a rank two symmetric tensor (a degenerate metric tensor)
\begin{equation}
\label{eq:g3-dv}
{}^{(3)}g = g_{\theta\theta}(\theta)d\theta^2 +  g_{\phi\phi}(\theta)\left(\beta_-d\phi +\left(\beta_+-\beta_-\right)dv\right)^2,
\end{equation}
and a $1$-form  
\begin{equation}
\label{omega3dv}
{}^{(3)}\omega = \omega_\theta(\theta) d\theta + \omega_\phi(\theta)(\beta_-d\phi +(\beta_+-\beta_-)dv). 
\end{equation}
Both are regular on this domain. 
On the other hand, on  $S^2\setminus \{p_-\}\times S^1$ consider the tensor
  \begin{equation}
  \label{eq:g3-dv'}
{}^{(3)}g = g_{\theta\theta}(\theta')d\theta'^2 +  g_{\phi\phi}(\theta')\left(\beta_+d\phi' +\left(\beta_--\beta_+\right)dv'\right)^2,    
\end{equation}
and a $1$-form
\begin{equation}
\label{omega3dv'}
{}^{(3)}\omega = \omega_\theta(\theta')d\theta' - \omega_\phi(\theta')(\beta_+d\phi' +(\beta_--\beta_+)dv'). 
\end{equation}
Again, both are regular on their domain.

Finally, consider $(\theta,\phi,v)$ and  $(\theta',\phi',v')$ as two charts related by the following transformation law:
\begin{equation}\label{transformation}
   \theta=\theta',
   \qquad
   \phi=-\phi',
   \qquad
   v=v'-\phi',
   \qquad\text{for}\qquad
   \theta,\theta'\neq 0,\pi.
\end{equation}
Denote the resulting $3$-dimensional manifold by $H$. It is equipped with a $C^{k+1}$ tensor ${}^{(3)}g$, a degenerate metric tensor, and a $C^k$ $1$-form ${}^{(3)}\omega$.  This structure admits  two symmetry generators:
\begin{equation}
  \xi_L = \partial_v = \partial _{v'},
  \qquad\text{and}\qquad
  \xi_R =  \partial_v+2 \partial_\phi = - \partial_{v'} -2\partial_{\phi'}.
\end{equation}
Each of them endows $H$ with the structure of a $U(1)$ principal fiber bundle over $S^2$. The null vector field  $\ell$ is 
\begin{equation}
\label{eq:ell-transversal-construction}
\ell=\frac{1}{\beta_+}\partial_v+\Big(\frac{1}{\beta_+}-\frac{1}{\beta_-}\Big)\partial_\phi =
\frac{1}{\beta_-}\partial_{v'}+\Big(\frac{1}{\beta_-}-\frac{1}{\beta_+}\Big)\partial_{\phi'}.
\end{equation}
If the following holds
$$\beta_-=\beta_+=\beta,$$ 
then \
$$\ell=\xi_L,$$ 
and the space $S$ of the null curves in $H$ is $S^2$, on which a metric tensor and differential $1$-form are induced, namely  
\begin{equation}
 g =  g_{\theta\theta}(\theta)d\theta^2 +  \beta g_{\phi\phi}(\theta) d\phi^2
\end{equation}
and a differential $1$-form 
\begin{equation}
\omega = \omega_\theta(\theta) d\theta - \beta \omega_\phi(\theta) d\phi^2. 
\end{equation}
such that 
\begin{align}
    {}^{(3)}g = \pi^*g,
    \qquad
    {}^{(3)}\omega = \pi^*\omega.
\end{align}
where 
\begin{equation}
    \pi : H\rightarrow \mathcal{S}
\end{equation}
is the natural projection. 
However, in the general case
\begin{equation}
\label{eq:ell-xis}
    \ell = \frac{1}{2(\beta_++\beta_-)}\left[
        \left(\beta_+ + \beta_-\right)\xi_L 
        + \left(\beta_+-\beta_-\right)\xi_R
    \right]
\end{equation}
and, generically, the space of the null curves is not a differentiable manifold, even though the structure of $H$ is regular.  
The genericity condition, implying that $\ell$ does not have closed orbits, is 
\begin{equation}
    \frac{\beta_+-\beta_-}{\beta_++\beta_-}\notin \mathbb{Q}. 
\end{equation}
 
A covariant derivative ${}^{(3)}\nabla$ is partially determined on $H$ by this data. 
Indeed, for a $1$-form $X$ such that
\begin{equation}
    \ell^aX_a = 0,   
\end{equation}
its covariant derivative is completely determined. Specifically, one has
\begin{equation}
    {}^{(3)}\nabla_aX_b = {}^{(3)}\nabla_{[a}X_{b]} + \frac{1}{2}{\cal L}_{\tilde{X}} {}^{(3)}g_{ab}, 
    \quad\text{where}\quad
    \tilde{X}^a {}^{(3)}g_{ab}= X_b,   
\end{equation}
and $\tilde{X}$ denotes an arbitrary vector field that satisfies the latter equality \cite{typeD}. To determine the remaining component of the covariant derivative, consider a $1$-form $n$, nowhere orthogonal to the null generators of $H$. We choose $n$ such that
\begin{align}
    \ell^an_a=-1,
\end{align}
in particular
\begin{align}
     n_a:=-\beta_+\nabla_a v,
\end{align} 
and define the tensor
\begin{align}
    S_{ab}:={}^{(3)}\nabla_a n_b.
\end{align}
From (\ref{nablaell}) it follows that
\begin{equation}
    \ell^a{}^{(3)}\nabla_bn_a = {}^{(3)}\omega_b.
\end{equation}
Hence, in the unprimed chart, the free part of  ${}^{(3)}\nabla$ is given by the pullback $S_{AB}$ of
\begin{align}
S_{ab} =-\beta_+{}^{(3)}\nabla_a{}^{(3)}\nabla_b v
\end{align}
to the surfaces $v=$const. Similarly, in the primed chart, the corresponding quantity is the pullback $S'_{AB}$ of 
\begin{align}
    S'_{ab}=-\beta_-{}^{(3)}\nabla_a{}^{(3)}\nabla_b v'
\end{align}
to the surfaces $v'=$const.
However, they have to satisfy the following constraint \cite{geometryhorizonsAshtekar_2002}:
\begin{align}\label{SprimS}
S_{AB} = \frac{\beta_+}{\beta_-}S'_{AB} +\beta_+ \nabla_A\nabla_B \phi'.
\end{align}
One of the pullbacks can be set arbitrarily, the other is determined via (\ref{SprimS}) in such a way that the connection ${}^{(3)}\nabla$ is defined globally on $H$.

Suppose finally, that the $2$-metric $g$ and the $1$-form $\omega$   satisfy the EEH equation (\ref{extremity}) on $S^2\setminus\{p_-,p_+\}$, namely
 \begin{equation}\label{extremity2}
  \nabla_{(A}\omega_{B)} + \omega_A\omega_B -\frac{1}{2}R_{AB} +\frac{\Lambda}{2}g_{AB} = 0.    
 \end{equation}
 Then also the pullbacks of ${}^{(3)}g$ and ${}^{(3)}\omega$ to any (local) spacelike section of $H$ satisfy (\ref{extremity2}),  hence, the resulting extremal horizon is vacuum.  

 Note that the above explicit construction following from the transformation law \eqref{transformation} endows the horizon with the topology of the Hopf bundle. 
 To construct the horizons with $L(n,1)$ lens space topology, it is enough to take instead the transformation:
 \begin{equation}\label{transformationN}
   \theta=\theta',
   \qquad
   \phi=-\phi',
   \qquad
   v=n v'-\phi',
   \qquad\text{for}\qquad
   \theta,\theta'\neq 0,\pi,
\end{equation}
for $n\in\mathbb{N}$.

Finally, we emphasize that if the conical singularities are regularizable ($\beta_+=\beta_-$) and the $U(1)-$fibers are null, the above construction differs from the one considered in \cite{hopf}.
The $U(1)-$ principal bundle structure of the horizons considered therein and also in \cite{LO4genus} used only objects naturally defined on any isolated horizon.
In particular, the rotation 1-form potential was used to introduce a $U(1)-$ bundle connection through a $u(1)-$valued connection 1-form
\begin{equation*}
    A:=\frac{{}^{(3)}\omega}{\kappa} \otimes \ell^*,
\end{equation*}
where $\ell^* \in u(1)$ is the Lie algebra element corresponding to the Killing vector field $\ell$.
However, in the extremal case, the three-dimensional 1-form ${}^{(3)}\omega$ on the horizon is entirely determined by -- and simply is the pullback of --  $\omega$ on $\mathcal{S}$. Consequently, it does not uniquely determine the connection 1-form.

\section{Solutions}
Instead of the fixed spherical coordinate system used in the previous section, let us assume coordinates $(x,\phi)$ such that the (conically singular at the poles) metric tensor takes the following form:
\begin{equation}
\label{gAxisSym}
    g = R^2 \left( \frac{dx^2}{P(x)^2} + P(x)^2 d\phi^2 \right). 
\end{equation}
The residual coordinate transformations are 
\begin{equation}\label{residual}
    x = x' - x_0,
    \quad
    \phi = \phi'
    \qquad\text{and}\qquad
    x = cx',
    \quad
    \phi = \frac{\phi'}{c}. 
\end{equation} 
The equation (\ref{extremity2}) admits an obvious symmetry 
\begin{equation}\label{ag}
    (g,\omega) \mapsto  (c^2g,\omega),
    \qquad
    c\in\mathbb{R}\setminus\{0\}
    . 
\end{equation}
When we are working with the coordinates used in $(\ref{g})$, the transformations (\ref{ag}) should be accompanied with the change of coordinate
\begin{equation}
    x' = cx,
    \quad
    \phi' = {c}\phi. 
\end{equation}

For a conically singular metric tensor, the Hodge dual  still defines a smooth  map 
\begin{equation}
    \star:\Omega^1\longrightarrow \Omega^1
\end{equation}
in the usual way.
Therefore, the rotation potential $\omega$, can be globally decomposed into an exact and co-exact part
\begin{equation}\label{BU}
    \omega = 
        \frac{dB}{B} + \star dU  =
        \frac{B_{,x}(x)}{B(x)}dx + P^2(x) U_{,x}(x)d\phi
    .   
\end{equation}
The metric tensor  (\ref{g}) defines a metric on $S^2$ with conical singularities at the poles $x_-$ and $x_+$, and smooth otherwise, provided $P$ is smooth on $[x_-,x_+]$ and all the following conditions hold:
\begin{align}
    &x \in [x_-,x_+], \\
    &P(x_-)=P(x_+) = 0, \label{smoothness}\\
    &P'(x_-),\ P'(x_+) \neq 0, \\ 
    &P(x) > 0 \quad\text{for}\quad x_- < x < x_+ \label{positivity}.
\end{align}    
For a function  $f:[x_-,x_+]\rightarrow \mathbb{R}$, to define a smooth function on the resulting $S^2$ it is sufficient to be smooth on $[x_-,x_+]$. 
Without loss of generality, we set $x_-=-1$, $x_+=1$ and, as a consequence of the possible conical singularity, $\phi \in [0;2\pi \beta)$, and adopt this choice in all subsequent calculations. 

The trace free part of the EEH equation (\ref{extremity2}) amounts to the following two equations:
\begin{equation}\label{UB}
	\left\{
	\begin{aligned}
		\partial_x^2B - B\left(\partial_xU\right)^2 &= 0 \\
		\partial_x\left(B^2 \partial_xU\right) &= 0
	\end{aligned}
	\right.
\end{equation}
In the following subsections, we derive rotating and non-rotating solutions for vanishing and non-vanishing cosmological constant $\Lambda$. The results are summarized in the Table \ref{table1}.

\subsection{Rotating $\Lambda\neq0$ solution}
\label{rotSol}

Let us first consider the case (see the latter equation above)
\begin{equation}
    B^2 \partial_xU = \text{const} \neq 0.     
\end{equation}
In this case 
\begin{equation}
   d\omega \not=0,
\end{equation}
and we call it rotating. 
The general solution of (\ref{UB}) for the functions $U$ and $B$ is 
\begin{align}
	B^2 &= B_0^2 \left[ \Omega^2 + (x - x_0)^2 \right], \\
	U &= \arctan\left( \frac{x - x_0}{\Omega} \right) + U_0.
\end{align}
The constants $U_0$ and $B_0$ represent ambiguity in the definition of the functions $B$ and $U$.  
Both $x_0$ and (non-vanishing) $\Omega$  are the integration constants.
The first of these can be given any value (including $x_0=0$) using the translation of the variable $x$.
Similarly, the second, as long as it is different from zero, can be varied arbitrarily using the scaling of $x$ (see (\ref{residual})). 
Notice that modulo those ambiguities,  the functions $U$ and $B$ are uniquely determined, and even independent of $\Lambda$.    

The trace part of (\ref{extremity2}) reads
\begin{equation} \label{eq:NHG_2}
    \partial_x^2 P^2 
    + \frac{2(x-x_0)}{(x-x_0)^2 + \Omega^2}\partial_x P^2 
    + \frac{4\Omega^2}{[(x-x_0)^2 + \Omega^2]^2} P^2 
    = -2\Lambda R^2,
\end{equation}
and it can also be easily solved. 
Assuming $\Lambda\neq0$, we find the general solution to be of the form 
\begin{equation}\label{generalP}
 P^2 = -\frac{\Lambda R^2(x-x_0)^2\Big((x-x_0)^2+3\Omega^2\Big)+3c_1\Big((x-x_0)^2-\Omega^2\Big)-6c_2(x-x_0)\Omega}{3\Big((x-x_0)^2+\Omega^2\Big)}.
\end{equation}
Going back to conditions (\ref{smoothness}) and setting $x_-=-1$, $x_+=1$, yields
\begin{align}\label{Pextr}
	P^2 = \frac{\Lambda}{3} R^2 \left(1-x^2\right) \frac{
		\left(\Omega ^2 \left(x^2-4 x x_0+6
   		x_0^2+1\right)+\left(x_0^2-1\right) (x-3 x_0) (x-x_0)+3 \Omega ^4\right)
   	}{
   		\left(x_0^2+\Omega ^2-1\right) \left((x-x_0)^2+\Omega ^2\right)
   	},
\end{align}
for $x_0^2+\Omega^2 \neq 1$. Otherwise, that is, when $x_0^2+\Omega^2 = 1$ in (\ref{generalP}), conditions (\ref{smoothness}) require the cosmological constant $\Lambda$ to vanish, leading to the same solution as in the case discussed in the next subsection.

Furthermore, we shall find the corresponding solutions to the rotation potential $\omega$. 
For the non-vanishing cosmological constant $\Lambda$ and $x_0^2+\Omega^2 \neq 1$, we have
\begin{align}\label{omegaBDLO}
    \omega &= \frac{x-x_0}{ (x-x_0)^2+\Omega^2} dx + P^2\frac{\Omega}{ (x-x_0)^2+\Omega^2}d\phi\nonumber\\
    &= \frac{x-x_0}{(x-x_0)^2+\Omega^2}dx+\frac{\Lambda}{3}R^2\Omega(1-x^2)\frac{(x-3x_0)(x-x_0)(x_0^2-1)+\Omega^2(x^2-4xx_0+6x_0^2+1)+3\Omega^4}{\Big((x-x_0)^2+\Omega^2\Big)^2(x_0^2-1+\Omega^2)}d\phi.
\end{align}
Next, define the rotation pseudo-scalar $\Xi$ via
\begin{align}
	d\omega =: \Xi\left( R^2 dx\wedge d\phi \right)
	\implies
	\Xi &= \frac{1}{R^2}\partial_x \left( P^2 \partial_x U\right),
\end{align}
which implies 
\begin{align}
\label{XiGen}
	\Xi &=-\frac{2 P^2\Omega(x - x_0)}{R^2\left[\Omega^2 + (x-x_0)^2 \right]^2}
	+\frac{2 PP_{,x}\Omega}{R^2[\Omega^2 - (x-x_0)^2]}\\
    &=\frac{2 \Omega}{R^2(\Omega^2 + (x-x_0)^2)}\bigg(-\frac{ P^2(x - x_0)}{\left(\Omega^2 + (x-x_0)^2 \right)}+ PP_{,x}\bigg)
\end{align}
For the rotation pseudo-scalar to be smooth at the poles, it has to satisfy 
\begin{equation}
	\left.\left(\left( \frac{d\Xi}{dx} \right)^2 P^2 \right)\right|_{x=\pm 1} = 0.
\end{equation}
Notice, however, that due to the properties of the frame coefficient $P^2$ and the form of the pseudo-scalar $\Xi$ the above condition is satisfied.

\subsection{Rotating $\Lambda=0$ solution}
    For the vanishing cosmological constant, solution (\ref{Pextr}) reduces to
    \begin{equation} \label{Pextr0}
        P^2 = \frac{
            c_1 \left(\Omega^2-(x-x_0)^2\right)
            + 2 c_2 \Omega  (x-x_0)
        }{
            \Omega^2 + (x-x_0)^2
        }
        .    
    \end{equation}
Applying conditions (\ref{smoothness}) leads to 
    \begin{equation}\label{labda0sol}
        P^2(x) = c_1\frac{(1-x^2)}{x^2- 2x_0 x+1},
    \end{equation}
    {where $c_1>0$ and $x_0^2 + \Omega^2 = 1$.}

It follows that the rotation 1-form potential $\omega$ is of the form
\begin{align}
    \omega  &= \frac{x-x_0}{ x^2-2xx_0+1} dx + \tilde c\frac{(1-x^2)\sqrt{1 - x_0^2}}{(x^2-2xx_0+1)^2}d\phi,
\end{align}
where $\tilde c= \pm c_1$, whereas the rotation pseudo scalar $\Xi$ in this case reads
\begin{align}
	\Xi =2\tilde c \frac{(x^3-3x+2x_0)\sqrt{1-x_0^2}}{R^2(x^2-2x x_0+1)^3}.
\end{align}

\subsection{Non-rotating  solution}
Finally, we go back to the case when
\begin{equation}
    B^2 \partial_xU =  0,    
\end{equation}
which, due to the first expression in  (\ref{UB}), leads to either $B=0$ (which we exclude since the rotation potential 1-form $\omega$ would be ill-defined) or
\begin{align}\label{nonrotsol0}
    U=\text{const}.
    \implies
    B(x) = a_1x+a_2.
\end{align}
For the potential $B$ to be well-defined, it must not vanish.
Without the loss of generality, we assume it to be positive everywhere and therefore $a_2 \geq |a_1|$. 
Notice that from \eqref{nonrotsol0} follows that
\begin{equation}
   d\omega =0, 
\end{equation}
and therefore we refer to it as a non-rotating solution. Trace part of (\ref{extremity2}) now reads 
\begin{align}\label{trace2}
    \partial_x^2P^2+\frac{2}{x+a_0}\partial_xP^2+2\Lambda R^2=0,
\end{align}
where $a_0:=a_2/a_1$ (assuming $a_1\neq0$ and $|a_0|>1$), and its solution is of the form 
\begin{align}
    P^2 = -\frac{1}{3}\Lambda R^2 x^2-\frac{2}{3}\Lambda a_0 R^2 x -\frac{\tilde c_1}{x+a_0} +\tilde c_2.
\end{align}
By applying boundary conditions (\ref{smoothness}) to the above we have
\begin{align}\label{specialSol}
    P^2 = \frac{\Lambda}{3} R^2 (1-x^2)\frac{x+3a_0}{x+a_0}.
\end{align}
Equation (\ref{trace2}) can be solved separately for vanishing parameter $a_1$, that yields
\begin{align}
    P^2 = -\Lambda R^2 x^2+ \hat c_1 x+\hat c_2.
\end{align}
Applying (\ref{smoothness}) yet again, gives
\begin{align}\label{specialspecial}
    P^2= \Lambda R^2 (1-x^2),
\end{align}
which, however, ends up being equivalent to (\ref{specialSol}) when taking a limit $a_0\rightarrow \pm\infty$ (or equivalently $a_1\rightarrow 0^\pm$).

The non-rotating solutions (\ref{specialSol}) and (\ref{specialspecial}) are degenerate for vanishing cosmological constant $\Lambda$. However, going back to equation (\ref{trace2}) and setting $\Lambda=0$, one can show there are no solutions satisfying conditions (\ref{smoothness}) for the frame coefficient $P^2$.

\section{Admissible parameter values}
In this section, we turn our attention to parameter constraints induced by the positivity condition of the frame coefficient $P^2$ (\ref{positivity}). The rotating solution for a non-vanishing cosmological constant can be written in the following form 
\begin{align}\label{P2recall}
	P^2  &= \frac{\Lambda}{3} R^2 \left(1-x^2\right) \frac{
		\left(\Omega ^2 \left(x-2  x_0\right)^2+\Omega ^2 \left(2 x_0^2+1\right)+\left(x_0^2-1\right) (x-3 x_0) (x-x_0)+3 \Omega ^4\right)
   	}{
   		\left(x_0^2+\Omega ^2-1\right) \left((x-x_0)^2+\Omega ^2\right)
   	},
\end{align}
where
\begin{align}
	\frac{1}{3} R^2 \left(1-x^2\right) \frac{1}{
   		 (x-x_0)^2+\Omega ^2
   	}>0.
\end{align}
Moreover, for $x_0^2\geq 1$ , we have
\begin{align}\label{ineq}
     W(x):=\frac{
		\Omega ^2 \left(x-2  x_0\right)^2+\Omega ^2 \left(2 x_0^2+1\right)+\left(x_0^2-1\right) (x-3 x_0) (x-x_0)+3 \Omega ^4
   	}{
   		\left(x_0^2+\Omega ^2-1\right)
   	}>0,
\end{align}
indicating that the positivity of $P^2$ in this case is determined by the positive cosmological constant. To conclude, if $x_0^2\geq 1$ (and $\Omega\neq0$), then the frame coefficient $P^2$ is positive if and only if $\Lambda>0$.

Therefore, we now move on to the case where $x_0^2< 1$ and investigate the value of the numerator of (\ref{P2recall}). The only term that could be negative is
\begin{align}\label{inequal}
    (x_0^2-1)(x-3x_0)(x-x_0)<0.
\end{align}
However, there is no value of $x_0\in(-1,1)$, for which the inequality (\ref{inequal}) holds for all $x\in(-1,1)$. Therefore, the numerator cannot be entirely negative either. We restrict our attention to solutions for which it is positive and divide the analysis into two subcases.

First, let us take
\begin{align}
    x_0^2+\Omega^2>1.
\end{align}
It follows that it is a sufficient condition for the numerator to be positive, and thus
\begin{align}
   \Big( x^2_0<1 \  \wedge  \ x_0^2+\Omega^2>1 \ \wedge \  \Lambda>0 \Big)  \implies P^2 > 0.
\end{align}
Notice, that together with the previous result for $x_0^2>1$ we can now conclude that in the case of $\Lambda>0$, the frame coefficient is positive whenever
\begin{align}
    x_0^2+\Omega^2>1.
\end{align}

\begin{figure}[ht!]
    \centering
    \includegraphics[width=0.75\linewidth]{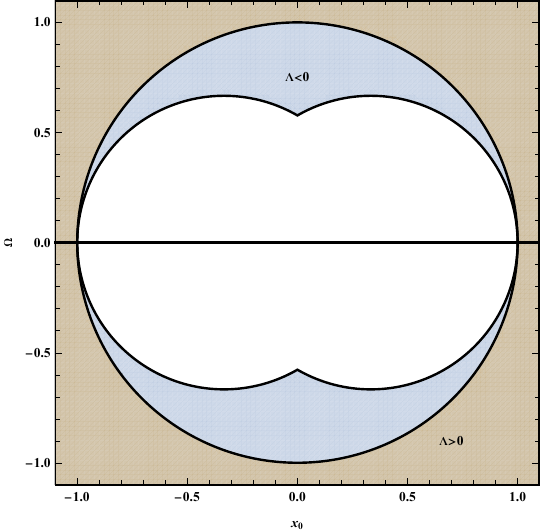}
    \caption{Admissible parameter values for the general rotating solution. For $\Lambda > 0$ all parameters outside of the outer circle are admissible. For $\Lambda < 0$ all values outside of the inner contour are admissible. We also exclude $\Omega\neq 0$.}
    \label{fig:parameters}
\end{figure}

Consider now the second sub-case, that is
\begin{align}
    x_0^2+\Omega^2<1,
\end{align}
 and study the second-order polynomial $W(x)$ which may be written in the form
\begin{align}
    W(x)= x^2-4x_0 x +4+3x_0^2+3\Omega^2-4\frac{x_0^2-1}{x_0^2-1+\Omega^2} .
\end{align}
Its discriminant reads
\begin{align}
    \Delta = 4\frac{(x_0^2-1-3\Omega^2)(x_0^2+\Omega^2)}{x_0^2-1+\Omega^2}.
\end{align} 
It is straightforward to check that for $x_0^2+\Omega^2<1$, it can only be positive. Therefore, the conditions for $W(x)$ to be positive are
\begin{align}
     &2x_0-\tfrac{1}{2}\sqrt{\Delta} \leq -1 \ \ \wedge  \ \ 2x_0+\tfrac{1}{2}\sqrt{\Delta} \geq 1.
\end{align}
Solving them yields
\begin{align}\label{conditionPneg}
    \Omega^2+\left(x_0-\tfrac{1}{3}\right)^2\geq\tfrac{4}{9} \ \ \wedge \ \ \Omega^2+\left(x_0+\tfrac{1}{3}\right)^2\geq\tfrac{4}{9}.
\end{align}
To conclude, for $\Lambda<0$ the frame coefficient $P^2$ is positive on the interval $(-1,1)$ if and only if (\ref{conditionPneg}) and $x_0^2+\Omega^2<1$ hold, as depicted in the Figure \ref{fig:parameters}.

For both the rotating solution with vanishing cosmological constant and the non-rotating solution, determining the admissible parameter values is straightforward, so we simply present the results in Table \ref{table1}.

{
\centering
\begin{table}[htbp]

\makebox[\textwidth]{\parbox{1.5\textwidth}{ %
\centering
\caption{{Summary of the horizon geometries obtained in this work.}}
\label{table1}
\[
\def\arraystretch{2.5}
\setlength{\arraycolsep}{6pt}
\begin{array}{|c|l|l|}
\hline
\textbf{\rotatebox[origin=c]{90}{\parbox{3cm}{\centering Solution\\ Type}}} 
& \multicolumn{1}{c|}{\textbf{Potentials and Frame Coefficient}}
& \multicolumn{1}{c|}{\textbf{Viable Parameters}} \\
\hline\hline
\multirow{4}{*}{\rotatebox[origin=c]{90}{\textbf{Rotating}}} 
& 
\begin{array}{l}
B^2 = B_0^2 \left[ \Omega^2 + (x - x_0)^2 \right] \\
U = \arctan\left( \frac{x - x_0}{\Omega} \right) + U_0 \\
P^2 = \frac{\Lambda}{3} R^2 \frac{
    (1 - x^2)\left(
    \Omega^2(x^2 - 4 x x_0 + 6 x_0^2 + 1)
    + (x_0^2 - 1)(x - 3x_0)(x - x_0)
    + 3 \Omega^4
\right)
}{
    (x_0^2 + \Omega^2 - 1)\left((x - x_0)^2 + \Omega^2\right)
} 
\end{array}
& 
\begin{array}{l}
\Lambda > 0,\ \Omega^2 + x_0^2 > 1,\ \Omega \neq 0 \\
\text{or} \\
\Lambda < 0,\ x_0^2 + \Omega^2 < 1,\\
\Omega^2+\left(x_0-\tfrac{1}{3}\right)^2\geq\tfrac{4}{9} \  \wedge \ \Omega^2+\left(x_0+\tfrac{1}{3}\right)^2\geq\tfrac{4}{9}
\end{array}
\\ \cline{2-3}

& 
\begin{array}{l}
B^2 = B_0^2 \left[ \Omega^2 + (x - x_0)^2 \right] \\
U = \arctan\left( \frac{x - x_0}{\Omega} \right) + U_0 \\
P^2 = c_1 \frac{1 - x^2}{x^2 - 2 x_0 x + 1}
\end{array}
& 
\begin{array}{l}
\Lambda = 0,\ c_1 > 0,\ \Omega \neq 0 \\
\Omega^2 + x_0^2 = 1
\end{array}
\\ \hline

\multirow{4}{*}{\rotatebox[origin=c]{90}{\textbf{Non-rotating}}} 
& 
\begin{array}{l}
B^2 = (a_1 x + a_2)^2  \\
U = \text{const.} \\
P^2 = \Lambda R^2 (1 - x^2) \frac{x + 3a_0}{3(x + a_0)}
\end{array}
& 
\begin{array}{l}
\Lambda > 0,\ a_0 \in (-\infty, -1) \cup (1, \infty) \\
a_0 = \frac{a_2}{a_1},\ |a_0|\geq 1
\end{array}
\\ \cline{2-3}

& 
\begin{array}{l}
B^2 = \text{const.} \\
U = \text{const.} \\
P^2 = \Lambda R^2 (1 - x^2)
\end{array}
& 
\begin{array}{l}
\Lambda > 0
\end{array}
\\ \hline
\end{array}
\]
}}
\end{table}
}

\section{Related solutions}

\subsection{Embedding into Plebański-Demiański spacetimes}
The natural candidates for the embedding spacetime for a horizon with a non-trivial bundle topology are Plebański-Demiański spacetimes with a non-vanishing NUT parameter.
The Plebański-Demiański metric is the most general solution to the ($\Lambda$-electro-) vacuum Einstein equations in four dimensions, such that the Weyl tensor is of Petrov type D \cite{PLEBANSKI197698}.
Its great physical importance comes from the fact that it includes as special limits the most commonly studied black hole solutions, among them the Schwarzschild and Kerr black hole metrics. 
We are interested in a subclass that is commonly used as a model of the exterior region of spherical black hole spacetimes.
The metric tensor for such spacetimes in the Griffiths-Podolský coordinates is given by \cite{gp2006}
\begin{equation}
\label{eq:KNdS-metric}
     {g}=\frac{1}{F^2}\left[-\frac{Q}{\Sigma}( {\dd}t-A   {\dd}\phi)^2  +\frac{\Sigma}{Q} {\dd}r^2 
+\frac{\Sigma}{\mathcal{P}} {\dd}\theta^2+\frac{\mathcal{P}}{\Sigma}\sin^2\theta(a {\dd}t-\rho  {\dd}\phi)^2\right]\;,
\end{equation}
 where we introduced the following functions:
\begin{equation}
\label{eq:metricFunctions-accKNadS}
    \begin{aligned}
        F(r,\theta)&=1-\frac{\alpha}{\bar \omega}r (l+a \cos{\theta})\;,\\
        \Sigma(r,\theta)&=r^2+(l+a\cos\theta)^2\;,\\
        \mathcal{P}(\theta)&=1-a_3\cos\theta-a_4\cos^2\theta\;,\\
        Q(r)&=\bar \omega^2k-2Mr+\epsilon r^2-2\frac{\alpha \nu}{\bar\omega}r^3-\left(\alpha k +\frac{\Lambda}{3}\right)r^4\;,\\
        A(\theta)&=\left(a\sin^2\theta-2l(\cos\theta-1)\right)\;,\\
        \rho(r)&=\Sigma+aA=\left(r^2+(a+l)^2\right)\;.\\
\end{aligned}
\end{equation} The constants appearing above are combinations of the spacetime parameters and are given by:
\begin{equation}
\begin{aligned}
        a_3 &= 2 \frac{\alpha a M} {\bar\omega} - 4 \alpha^2 \frac{a l}{\bar\omega^2} k\left( \bar\omega^2k+e^2+g^2\right) -\frac{4}{3} \Lambda a l\;,\\
        a_4 &= -\alpha^2 \frac{a^2}{\bar\omega^2}\left(\bar\omega^2+e^2+g^2\right) k - \frac{1}{3}\Lambda a^2\;,\\
        \epsilon&= \frac{\bar\omega^2 k}{a^2 - l^2} + 4 \frac{\alpha l M }{\bar\omega}-(a^2+3l^2)\left(\frac{\alpha^2}{\bar\omega^2} \left(\bar\omega^2k+e^2+g^2\right) + \frac{\Lambda}{3}\right)\;,\\
        n& = \frac{\bar\omega^2 k l}{a^2 - l^2} - \frac{\alpha M (a^2 - l^2)}{\bar\omega} + (a^2 - l^2) \left(\frac{\alpha^2}{\bar\omega^2} \left(\bar\omega^2k+e^2+g^2\right) + \frac{\Lambda}{3}\right)\;,\\
        k&=\frac{1 + 2 \alpha l M\bar\omega^{-1} -3\alpha^2l^2\bar\omega^{-2}\left(e^2+g^2\right)- l^2 \Lambda}{3\alpha^2l^2 + \bar\omega^2 {(a^2-l^2)}^{-1}}\;.
    \end{aligned}
\end{equation}

By the assumption on the Lorentzian signature of the metric tensor \eqref{eq:KNdS-metric}, we have $\mathcal{P}>0$.
It may be shown that the vanishing of $\Sigma$ and $F$ corresponds to a curvature singularity and the conformal infinity \cite{Griffiths_2005,gp2006}.
On the other hand, the zeros of $\mathcal{Q}$ are regular and correspond to Killing horizons.
If the cosmological constant is present, there are up to 4 distinct horizons, and in certain sub-cases, two of them may be interpreted as black hole horizons, while the other two as cosmological/acceleration horizons \cite{PodolskyVratny2022}.
However, in general, the causal properties of the horizons have to be established case by case.

In total, the family is described by 7 real numbers: $\alpha$, $a$, $l$, $m$, $e$, $g$, and $\Lambda$, which we will call the acceleration, Kerr, NUT, and mass parameters, electro-magnetic charges, and the cosmological constant.
The remaining parameter $\bar\omega$ is a gauge parameter and may be set to a non-zero value.
Convenient values for the gauge parameter $\bar\omega$ have been proposed, e.g. $\bar\omega=(a^2+l^2)/a$. 
However, because it requires a different parametrization \cite{PodolskyVratny2022}, we opt to keep $\bar\omega$ for the full generality.
The interpretation of $e$, $g$, and $\Lambda$ comes directly from the ($\Lambda$-)vacuum Einstein-Maxwell equations, which are satisfied by those spacetimes.
In certain limits, when one or more parameters vanish, $\alpha$, $m$, and $a$ are interpreted as the acceleration, mass, and the angular momentum per unit mass of a black hole.
However, importantly, in the most generic case, the physical meaning of the parameters is not clear.
Finally, the physical interpretation of the NUT parameter is not settled.
Depending on the interpretation of the spacetime, it gives rise to one of two types of topological singularity.

In the approaches following Bonnor's work \cite{bonnor_1969}, the axis of rotation is a quasi-regular, torsion singularity and may be further interpreted as a source of angular momentum.
This type of singularity is often referred to as a Misner string and is characterized by two values: conicity, measuring the failure of the axial Killing vector field to be $2\pi$ periodic, and the time-shift, measuring the failure of the axial Killing vector field to have closed orbits 
\footnote{We follow the convention that a cyclic Killing vector field has closed orbits, while the axial Killing vector fields have fixed points, which are referred to as the axis of rotation.
In generic spacetimes, they are not equal even up to normalization \cite{ConicalsingularityinspacetimeswithNUT}.}.
As was recently shown, if the NUT parameter is present, the values of the conicity and time-shift are observer-dependent, in particular, there exist observers for whom the conicity vanishes \cite{ConicalsingularityinspacetimeswithNUT}.

The alternative approach to the NUT spacetimes, which is crucial to find the embeddings of our solutions, was proposed by Misner \cite{misner}.
His idea was to change the topology of the surfaces of constant $r$ by promoting a timelike Killing vector field to a cyclic one.
With this choice, the coordinates $(t,\theta,\phi)$ of \eqref{eq:KNdS-metric} become the Euler angles and the surface $r={\text{const}}$ has the topology of a 3-sphere.
Additionally, any such surface, including the horizons, admits the structure of a Hopf bundle or other $U(1)$-principal fiber bundle over $S^2$.
In turn, this endows the spacetime with the structure of a $U(1)$ principal fiber bundle over $S^2\times \mathbb{R}$.
This procedure has been extended to the broader family of Plebański-Demiański spacetimes.
To achieve that, it is necessary to carefully choose the Killing field that will be promoted to the cyclic symmetry; in particular, the cyclic field does not have to coincide with the Killing vector field generating the horizon.
It turns out that if the NUT parameter is present, exactly two choices of the Killing vector fields exist such that it is possible to impose the $U(1)-$principal fiber bundle structure and the Misner interpretation \cite{LO3PhysRevD.104.024022,DLO}.
Recently, it was shown that those two choices exactly correspond to the observers for whom the difference of the conicities measures for $\theta=0$ and $\theta=\pi$ parts of the axis vanishes \cite{ConicalsingularityinspacetimeswithNUT}.
The explicit formulas for those Killing vector fields may be found in \cite{DLO}, in Section V.

Incidentally, the two interpretations of the NUT parameter lead to two possible types of horizons one could construct using solutions to the EEH equation on a sphere with conical singularities.
The first type is a horizon with a product topology, and the space of null generators being a two-sphere with conical singularities, similarly to Bonnor's interpretation, where the Misner string introduces a singularity of the axis.
However, by following the procedure outlined in Section \ref{sec:ExtremalNUTIsolatedHorizons} a novel type of extremal horizons with Hopf bundle structure may be constructed.

As the next step, we seek the explicit relation between the geometries and parameters of the abstract, extremal horizon with the metric tensor of the form \eqref{gAxisSym} and the frame function $P^2$ satisfying the trace part of \eqref{extremity2}, and on the other hand, the embedded extremal Killing horizons of the Plebański-Demiański spacetime.
To obtain the Riemannian metric on the space of the null generators of a horizon, one needs to rewrite the metric \eqref{eq:KNdS-metric} in coordinates covering the horizon and restrict it to the horizon hypersurface.
This procedure has been detailed in \cite{DLO}.
It follows that the necessary comparison is
\begin{align}
\label{eq:comp}
R^2\bigg( \frac{1}{P^2}dx^2+ P^2 {\color{black} }\beta^2 d\varphi^2 \bigg)=\frac{1}{F_H^2}\bigg( \frac{\Sigma_H}{\mathcal{P}}d\theta^2+\frac{\mathcal{P}}{\Sigma_H}\rho_H^2\sin^2\theta \beta^2 d\varphi^2\bigg),
\end{align}
where $F_H$, $\Sigma_H$ and $\rho_H$ are restrictions of the corresponding functions appearing in \eqref{eq:KNdS-metric} to the horizon of radius $r_H$.
The cyclic coordinates have been rescaled such that $\varphi$ is $2\pi$-periodic and $\beta$ is shorthand for $\beta_+$ or $\beta_-$, depending on whether the map smoothly covers the pole $x=1$ or $x=-1$ \cite{DLO}.

Subsequently, following the comparison procedure, we have \cite{DLO}
\begin{align}\label{x=cos}
x = \frac{\cos\theta(\bar\omega-\alpha lr_H)-a\alpha r_H}{a\alpha r_H\cos\theta+\alpha l r_H-\bar\omega},
 \end{align}
or equivalently
\begin{align}\label{cos=x}
\cos\theta=\frac{a\alpha r_H+(\alpha lr-\bar \omega)x}{\bar \omega-\alpha lr_H-a \alpha r_H x}.
\end{align}
Notice that for these transformations to be well defined, the denominators on the right-hand side of the above equations cannot vanish; therefore, the parameters must satisfy the following constraint:
\begin{equation} \label{eq:transformation_constraint}
    \left| \frac{\alpha r_H a}{\bar\omega - \alpha l r_H} \right| < 1
    .
\end{equation}
This inequality indeed holds.
The area of a horizon in Plebański-Demiański spacetime has been calculated in \cite{DLO}, and its radius (squared) is
\begin{equation}
    R^2 = \frac{
        \left[r_H^2 + (a+l)^2\right]\bar\omega^2
    }{
    \left[\bar\omega + \alpha r_{H}(a-l)\right]
    \left[\bar\omega - \alpha r_{H}(a+l)\right]
    }.
\end{equation}
For $R^2$ to be well defined, there has to be
\begin{equation}
    \left[\bar\omega + \alpha  r_{H}(a-l)\right]
    \left[\bar\omega - \alpha  r_{H}(a+l)\right] > 0,
\end{equation}
which is equivalent to (\ref{eq:transformation_constraint}).
Importantly, for non-accelerating spacetimes, the relations (\ref{x=cos})--(\ref{cos=x}) simplify to
\begin{equation*}
    x=-\cos\theta.
\end{equation*}
Finally, the frame function written explicitly in terms of $\theta$ reads
\begin{align}
P^2&=\frac{\mathcal{P}}{F^2\Sigma_H\xi}\rho_H\sin^2\theta \nonumber\\
&=\frac{1-a_3\cos\theta-a_4\cos^2\theta}{(1-\frac{\alpha}{\bar \omega}(l+a\cos\theta)r_H)^2( r_H^2+(l+a\cos\theta)^2)\xi}\rho_H\sin^2\theta,
\end{align}
where 
\begin{equation*}
    \xi=\frac{\bar \omega^2}{(\bar \omega+r_H\alpha(a-l))(\bar \omega-r_H\alpha(a+l))}    
\end{equation*}
is a convenient, constant combination of parameters.

The extremality condition for Plebański-Demiański spacetimes is simply
\begin{equation}
    \mathcal{Q}'(r)\eval_{r=r_H}=0,
\end{equation}
where $r_H$ is the radius of a horizon, i.e. it satisfies $\mathcal{Q}(r_H)=0$.
As the function $\mathcal{Q}$ is generically a fourth-degree polynomial, the explicit formulas for its roots, although attainable, are neither insightful nor easy to manipulate.
Instead, we opt to determine the mass parameter $m$,  which appears linearly in $\mathcal{Q}$, corresponding to the horizon.
This significantly simplifies the calculations.
However, because we don't solve $\mathcal{Q}(r)=0$ explicitly, some information is lost. We do not know which horizon, relative to the others, is investigated and how many other horizons are present.
A given value of $m$ satisfying $\mathcal{Q}=0$ may correspond to as many as 4 different values of  $r_H$.

\subsubsection{Embedding of rotating solutions}
We expect the generic rotating solutions to be embeddable in spacetimes with a non-vanishing Kerr parameter.
This intuition is made precise by investigating the rotation 1-form of the embedded horizons, calculated explicitly in \cite{DLO}.
In particular, the formula for the $\omega_\phi$ component, the only possibly contributing to $\dd \omega$, reads
\begin{equation}
\label{eq:omega_phi_embedded_hor}
    \omega_{\phi} =\frac{ a \rho_H r_H \mathcal{P}\sin^2\theta}{\Sigma_H^2}-\kappa A \frac{ \rho_H}{\Sigma_H}.
\end{equation}
It follows from the explicit calculation, given $\rho\,\mathcal{P}\,\Sigma\neq0$, that for $\kappa=0$ the only possibilities of $\partial_\theta \omega_{\phi}=0$ are given either {by vanishing $r_H$ or a}.
For now, we focus on the $a\,r_H\neq 0$ case.
The possibility of $a=0$ or $r_H=0$ will be investigated later, together with the embeddings of non-rotating horizons.

If the cosmological constant is present, then the extremality condition for a horizon of radius $r_H$, is solved explicitly by 
\begin{equation}
\begin{split}
    \Lambda&=\Big[3 \xi^2 \bar\omega ^2 \left(-a^2+l^2+r_H^2\right)\Big]\\
    &\times \Big\{\alpha  a^4 r_H \left(3
   \alpha  l^2 r_H-2 l \omega +\alpha  r_H^3\right)-a^2 \left[-8 \alpha 
   l^3 r_H \bar\omega +\bar\omega ^2 \left(3 l^2-r_H^2\right)+\alpha ^2
   r_H^2 \left(6 l^4+3 l^2 r_H^2+r_H^4\right)\right]\\
   &+3
   \left(l^2+r_H^2\right)^2 (\bar\omega -\alpha  l r_H)^2\Big\}^{-1},\\
    m&=\Big\{r_H \bar\omega  \left(a^4 (3 \alpha  l r_H+\bar\omega )+2 a^2 \left(-3
   \alpha  l^3 r_H+l^2 \bar\omega +\alpha  l r_H^3+r_H^2 \bar\omega
   \right)+\left(3 l^2-r_H^2\right) \left(l^2+r_H^2\right) (\alpha  l
   r_H-\bar\omega )\right)\Big\}\\
   &\times \Big\{\alpha  a^4 r_H \left(3
   \alpha  l^2 r_H-2 l \bar\omega +\alpha  r_H^3\right)-a^2 \left[-8 \alpha 
   l^3 r_H \bar\omega +\bar\omega ^2 \left(3 l^2-r_H^2\right)+\alpha ^2
   r_H^2 \left(6 l^4+3 l^2 r_H^2+r_H^4\right)\right]\\
   &+3
   \left(l^2+r_H^2\right)^2 (\bar\omega -\alpha  l r_H)^2\Big\}^{-1}.
   \end{split}
\end{equation}
Comparing the functions $P^2$ with the rotating $\Lambda\neq0$ solution \eqref{generalP} we get
\begin{equation}
    \begin{split}
    x_0&=\frac{\alpha  r_H \bar\omega  \left(a^2+r_H^2\right)-\alpha  l^2
   r_H \bar\omega +l \left(\bar\omega ^2-\alpha ^2 r_H^4\right)}{a
   \left(\alpha ^2 r_H^4+\bar\omega ^2\right)},\\
   \Omega^2&=\frac{r_H^2 \left((\bar\omega -\alpha  l r_H)^2-a^2 \alpha ^2
   r_H^2\right)^2}{a^2 \left(\alpha ^2 r_H^4+\bar\omega ^2\right)^2}.
    \end{split}
\end{equation}
For convenience, we also give the result for a sub-case of $\alpha=0$, i.e. non-accelerating Kerr-NUT-(a)dS spacetimes.
Then the extremality condition simplifies to
\begin{equation}
\begin{split}
    \Lambda&=\frac{3 \left(a^2-l^2-r_H^2\right)}{a^2 \left(3 l^2-r_H^2\right)-3
   \left(l^2+r_H^2\right)^2},\\
   m&=\frac{r_H \left(a^4+2 a^2 \left(l^2+r_H^2\right)-3 l^4-2 l^2
   r_H^2+r_H^4\right)}{a^2 \left(r_H^2-3 l^2\right)+3
   \left(l^2+r_H^2\right)^2},
   \end{split}
\end{equation}
and the parameter relations simplify to
\begin{equation}
    \begin{split}
        x_0&=\frac{l}{a},\\
        \Omega^2&=\frac{r_H^2}{a^2}.
    \end{split}
\end{equation}

By the same argument as for the non-vanishing $\Lambda$, the rotating solutions with $\Lambda=0$ are expected to be embeddable in accelerated Kerr-NUT spacetimes.
The extremality condition is expressed as
\begin{equation}
\begin{split}
    m&=\frac{\bar\omega  \left(a^2 (2 \alpha  l r_H+\bar\omega )-2 \alpha  l^3 r_H-l^2
   \bar\omega +r_H^2 \bar\omega \right)}{2 (\alpha  l r_H-\bar\omega ) \left(a^2
   \alpha  l-\alpha  l^3-r_H \bar\omega \right)},\\
   r_H&=\sqrt{a^2-l^2}.
   \end{split}
\end{equation}
The comparison with the rotating $\Lambda=0$ solution \eqref{labda0sol} yields
\begin{equation}
\begin{split}
    x_0&=\frac{2 \alpha \bar\omega (a^2-l^2)^{3/2}  +l \left(\bar\omega ^2-\alpha ^2
   \left(a^2-l^2\right)^2\right)}{a \left(\alpha ^2
   \left(a^2-l^2\right)^2+\bar\omega ^2\right)},\\
   c_1&=\Big\{2 (a+l) \big[\alpha ^6 a^{10} l^2-2 \alpha ^2 a^4 \left(5 \alpha ^4 l^8+4
   \alpha ^2 l^4 \bar\omega ^2+\bar\omega ^4\right)+\alpha ^2 a^2 l^2 \left(\alpha ^2
   l^4+\bar\omega ^2\right) \left(4 \alpha  l \bar\omega  \sqrt{a^2-l^2}+5 \alpha ^2
   l^4+\bar\omega ^2\right)\\
   &-\left(\alpha ^2 l^4+\bar\omega ^2\right)^2 \left(2 \alpha  l
   \bar\omega  \sqrt{a^2-l^2}+\alpha ^2 l^4-\bar\omega ^2\right)+\alpha ^4 a^8 \left(2
   \alpha  l \bar\omega  \sqrt{a^2-l^2}-5 \alpha ^2 l^4+\bar\omega ^2\right)\\
   &+2 \alpha ^4
   a^6 l^2 \left(-2 \alpha  l \bar\omega  \sqrt{a^2-l^2}+5 \alpha ^2 l^4+\bar\omega
   ^2\right)\big]\Big\}\Big\{a \left(\alpha ^2 l^2 (l-a) (a+l)+\bar\omega ^2\right)^2
   \left(\alpha ^2 \left(a^2-l^2\right)^2+\bar\omega ^2\right)\Big\}^{-1}.
\end{split}
\end{equation}

Additionally, in the special case when the acceleration parameter and the cosmological constant both vanish, $\mathcal{Q}$ reduces to a quadratic polynomial, and the explicit values of $r_H$ may be easily obtained.
The extremality conditions solve to $r_H=m=\pm\sqrt{a^2-l^2}$. 
The horizons are embeddable into Kerr-NUT spacetime with
\begin{equation}
        c_1=2\frac{a+l}{a},\quad x_0=\frac{l}{a}.
\end{equation}

\subsubsection{Embedding of non-rotating solution}
As argued at the beginning of the previous section, the non-rotating horizons are expected to be embeddable in Plebański-Demiański spacetimes without the Kerr parameter.
Such spacetimes are often called accelerating Taub-NUT-(anti-) de Sitter.
Interestingly, they are not covered by the coordinates and parametrization given by the generic accelerating Kerr-NUT-(anti-) de Sitter metric tensor \eqref{eq:KNdS-metric}.
Their existence within the Plebański-Demiański family has been established only recently, first without the cosmological constant \cite{Astorino2024_1}, and later with its inclusion \cite{Astorino2024_2}.
These metrics have been subsequently identified within the larger Plebański-Demiański family and written in Griffiths-Podolský-like coordinates \cite{PodolskyAstorino2025_1,PodolskyAstorino2025_2}, similar to the coordinates in \eqref{eq:KNdS-metric}.
Unfortunately, unless the cosmological constant also vanishes in addition to the Kerr parameter, the other parameters of this solution are given in implicit form, preventing a straightforward relation between the horizon and spacetime parameters.
Therefore, we refrain from obtaining the explicit relation between the parameters.

A non-rotating horizon within the accelerated Kerr-NUT-(anti-) de Sitter spacetime may be located at $r_H=0$, as argued from \eqref{eq:omega_phi_embedded_hor}.
Such a horizon is extremal if the parameters additionally satisfy the following constraints
\begin{equation}
    m=0,\quad \Lambda=\frac{1}{l^2}.
\end{equation}
Note that even though $m=0$, this spacetime is not simply de Sitter as the NUT parameter contributes to the curvature. 
Then, at $r_H=0$ we have $F_H=1$ and $\xi=1$ and so the horizon frame coefficient simplifies to 
\begin{equation}
    P^2=\frac{\left(1-x^2\right) (a+l)^2 (3 l- a x)}{3 l^2 (l-a x)},
\end{equation}
and the direct comparison yields
\begin{equation}
    \begin{split}
        R^2&=(a+l)^2,\\
        a_0&=-\frac{l}{a}.
    \end{split}
\end{equation}

Finally, the last solution from the Table \ref{table1} has the same conicities at both $x=\pm1$.
Therefore, the conical singularity of such a horizon is regularizable at both poles simultaneously by rescaling the cyclic coordinate.
Such a horizon is expected to be embeddable in Taub-NUT-de Sitter spacetime, which is confirmed by direct calculation.
The comparison results in
\begin{equation}
    R^2=r_H^2+l^2.
\end{equation}

\subsection{Solutions to the Petrov type D equation}
Horizons of the same non-trivial, Hopf bundle structure with null curves transversal to the bundle fibers satisfying the Petrov type D equation
\begin{align}\label{typeD}
    \bar m^A \bar m^B \nabla_A \nabla_B\left( K-\tfrac{\Lambda}{3}+i\Xi\right)^{-\frac{1}{3}}=0
\end{align}
have been derived in \cite{DLO}. Eq. (\ref{typeD}) is an integrability condition for the EEH equation (\ref{extremity}), meaning that our extremal horizons could be identified among the solutions of the type D equation. Nevertheless, the Petrov type of the spacetime Weyl tensor does not have to, and generically will not be, of the type D. The type D equation is the constraint strictly for non-extremal isolated horizons to be of the Petrov type D. The frame coefficient $P^2$, which is the solution to the real part of eq. (\ref{typeD}) generically is of the form
\begin{align}\label{Pfinal}
P^2&=R^2\frac{-\frac{\Lambda}{3}x^2+Cx+D}{x^2+A x+B}(x^2-1).
\end{align}
The comparison with the frame coefficient for the extremal case (\ref{Pextr}) with a non-vanishing cosmological constant yields
\begin{align}
    A&=-2x_0, \\
    B&=x_0^2 +\Omega^2, \\
    C&=\frac{4}{3}\Lambda x_0, \\
    D&=-\frac{\Lambda}{3} \ \frac{ 3 x_0^4 + \Omega^2 + 
    3 \Omega^4 + x_0^2 (-3 + 6 \Omega^2)}{
  -1 + x_0^2 + \Omega^2}.
\end{align}
That means that we may write the generic expression (\ref{Pextr}) in terms of the parameters $A$ and $B$, that is
\begin{align}
    P^2= R^2\frac{\Lambda}{3}\frac{x^2+2Ax-\frac{A^2-B(1+3B)}{B-1}}{x(x+A)+B}(1-x^2)
\end{align}
and parameters $A$ and $B$ must be such, that $P^2$ is positive everywhere on $(-1,1)$, that is
\begin{align}\label{P^2>01}
    \frac{x^2+Ax-\frac{A^2-B(1+3B)}{B-1}}{x(x+A)+B}>0 \ \ \ \text{for} \ \ \ \Lambda>0, \\\label{P^2>02}
    \frac{x^2+Ax-\frac{A^2-B(1+3B)}{B-1}}{x(x+A)+B}<0 \ \ \ \text{for} \ \ \ \Lambda<0.
\end{align}
Similarly, we compare the solution to the imaginary part of the type D equation with our result, that is, the rotation pseudoscalar (\ref{XiGen}) for $\Lambda\neq 0$, which yields
\begin{align}
    \Xi=\frac{\Lambda}{3} \ \frac{\sqrt{4B-A^2}\left( A^3+(1+B)^2(x^2-3B)x -A\left(1+B\right)\left(2B+3\left(B-1\right)x^2\right)+A^2\left( 3-Bx^2\right)x \right)}{\left(1-B\right)\left(B+x\left(A+x\right)\right)^3}.
\end{align}
The above is well defined whenever:
\begin{align}
    4B-A^2>0
     \ \Longleftrightarrow \ 
    4\Omega^2 > 0,
\end{align}
which implies that both of the denominators in (\ref{P^2>01}) and (\ref{P^2>02}) are always positive. This simplifies the positivity requirement for the frame coefficient $P^2$, namely:
\begin{align}\label{P^2>01v2}
    {x^2+Ax-\frac{A^2-B(1+3B)}{B-1}}>0 \ \ \ \text{for} \ \ \ \Lambda>0, \\\label{P^2>02v2}
    {x^2+Ax-\frac{A^2-B(1+3B)}{B-1}}<0 \ \ \ \text{for} \ \ \ \Lambda<0.
\end{align}

Next, we investigate the case of vanishing cosmological constant, that is, solution (\ref{labda0sol}), which gives the following relations between the parameters $A$, $B$, $C$, $D$ and $c_1$, $x_0$:
\begin{align}
    A&=-2x_0,\\
    B&=1,\\
    C&=0,\\
    R^2D&=-c_1,
\end{align}
whereas the rotation pseudoscalar, up to the overall sign, is of the form:
\begin{align}
    \Xi = \frac{\sqrt{4-A^2}D(x^3-3x-A)}{(1+x(A+x))^3}.
\end{align}
Therefore, we have also identified the solutions of the EEH equation with cosmological constant (\ref{extremity}) among the solutions of the Petrov type D equation (\ref{typeD}) of the same bundle structure.

\subsection{Lucietti-Kunduri solutions}

A comprehensive review of solutions to the extremal horizon equation, incorporating various spacetime dimensions, horizon topologies, and symmetry assumptions, is presented in \cite{Lucietti2013}. In particular, extremal horizons invariant under a $\mathbb{R} \times U(1)$ isometry have been examined. The corresponding solution—originally derived in \cite{Hajicek, Li_2013} for $\Lambda = 0$, extended in \cite{lucietti2009} to the $\Lambda < 0$ case, and later generalized in \cite{Lucietti2013} to arbitrary values of the cosmological constant—takes the form \footnote{Note that, to avoid a clash of notation, we do not use the same symbols as in \cite{Lucietti2013} to denote the function and parameter appearing in the metric g and 1-form $\omega$. Specifically, the function $\mathcal{B}$ and parameter $\gamma$ correspond used here correspond to $\mathcal{P}$ and $\beta$, respectively, in \cite{Lucietti2013}.}
\begin{align}\label{gLK}
    g &=\frac{dx^2}{B
    (x)}+B(x)d\phi^2,\\\label{omegaLK}
    \omega &= \frac{\Gamma'}{2\Gamma}dx-\frac{kB(x)}{2\Gamma}d\phi,
\end{align}
where 
\begin{align}
    B&= \frac{\mathcal{B}(x)}{\Gamma},\\
    \mathcal{B}(x)&=-\frac{\gamma\Lambda}{12}x^4+ \left(A_0-2\Lambda k^2\gamma^{-1}\right)x^2+\hat c_1 x -\frac{4k^2}{\gamma^2}\left(A_0-\Lambda k^2\gamma^{-1}\right),\\
    \Gamma&={\frac{k^2}{\gamma}+\frac{\gamma x^2}{4}}.
\end{align}
Notice that rescaling all parameters by $\gamma$, namely
\begin{align}
    \tilde A_0 := A_0/\gamma, && \tilde k := k/\gamma, && \tilde c_1 := \hat c_1/\gamma,
\end{align}
reduces the solution to
\begin{align}
    \mathcal{B}(x)&=\gamma\left(-\frac{\Lambda}{12}x^4+ \left(\tilde A_0-2\Lambda \tilde k^2\right)x^2+\tilde c_1 x -{4\tilde k^2}\left(\tilde A_0-\Lambda \tilde k^2\right)\right),\\
    \Gamma&=\gamma\left({\tilde k^2+\frac{ x^2}{4}} \right).
\end{align}
Now the frame coefficient simplifies to
\begin{align}
    B=\frac{-\frac{\Lambda}{3}x^4+ 4\left(\tilde A_0-2\Lambda \tilde k^2\right)x^2+4\tilde c_1 x -{16\tilde k^2}\left(\tilde A_0-\Lambda \tilde k^2\right)}{4\tilde k^2+ x^2}.
\end{align}
This solution was derived purely locally, without imposing any assumptions on the topology of the horizon. Therefore, we expect it to coincide with our general solution~(\ref{generalP}). Before comparing the metric coefficients, we first rewrite the metric~(\ref{gAxisSym}) in the same form as~(\ref{gLK}). This can be achieved by introducing the coordinate transformation $\tilde x = R^2 x$ in~(\ref{gAxisSym}), which yields:
\begin{align}
    g&= \frac{ d\tilde x^2}{R^2P^2}+R^2P^2d\phi^2.
\end{align}
By direct comparison of the component $g_{\phi\phi}$, namely $B$ and $R^2P^2$, we find the following relations between Lucietti-Kunduri parameters and parameters $\dot \Omega:=R^2\Omega$, $\dot c_1:=R^2 c_1$ and $\dot c_2:=R^2 c_2$
\begin{align}
    \tilde k^2&=\tfrac{1}{4}\dot\Omega^2,\\
    \tilde A_0&=\tfrac{1}{4}\left(\Lambda\dot\Omega^2-\dot c_1  \right),\\
    \tilde c_1&=\tfrac{1}{2}\dot c_2 \dot\Omega
\end{align}
and due to local freedom in Lucietti-Kunduri solution $x_0=0$. We find that the rotation 1-form potential \eqref{omegaLK}, when written in terms of parameters $\Omega$ and $c_1$ (with $x_0=0$) reproduces the expression in the first line of \eqref{omegaBDLO}, i.e., before inserting the explicit form of the frame coefficient $P^2$.
Therefore, it follows that there is a one-to-one correspondence between these solutions. This means that taking any solution given by Lucietti-Kunduri and imposing conditions \eqref{smoothness} on the poles results in the rotating $\Lambda\neq0$ solution described in Section \ref{rotSol}.

\subsection{Podolský-Matejov solutions}
Recently, the extremal isolated horizons have been investigated by means of the Newman-Penrose formalism. The Newman-Penrose equations and Bianchi identities were analyzed on extremal isolated horizons and solved first for vanishing cosmological constant \cite{PodolskyMatejov2021}, and subsequently with its inclusion \cite{PodolskyMatejov2022}. These equations are equivalent to the EEH equation \eqref{extremity2} for the vanishing electromagnetic field considered in this work. The authors have also found the embedding spacetimes, whenever it was possible to explicitly solve for the horizon radius.

First, we note that the parameters $\beta_\pm$ are the inverse of the conicities $\mathcal{C}_\pm$ of the horizon axis at $\theta=0$ and $\theta=\pi$, respectively.
This is most easily seen by following the circumference-radius definition of the conicity, i.e. 
\begin{equation}
    \mathcal{C}=\lim_{x \to \text{axis}} \frac{\text{circumference}}{2\pi\ \text{radius}} ,
\end{equation}
where the circumference is calculated along the orbit of cyclic symmetry, the radius is geodesic and $x$ parametrizes the closed orbits.
This calculation for the metric tensor \eqref{gAxisSym} gives
\begin{equation}
\mathcal{C}_{\pm}:=\frac{1}{2}\partial_x(P(x)^2)\eval_{x=\pm1}.
\end{equation}
Regular axis without the conical singularity corresponds to $\beta_\pm=1$ or equivalently $\mathcal{C}_\pm=1$, as we have 
\begin{equation}
    \mathcal{C}_\pm=\frac{1}{\beta_\pm}.
\end{equation}
The $\beta$ parameters for the rotating $\Lambda\neq0$, rotating $\Lambda=0$ and non-rotating solution read
\begin{align}
         \beta_\pm&=\Lambda  R^2 \pm \frac{2 \Lambda  R^2 (x_0 \pm 1)}{3
   \left(x_0^2+\Omega^2-1\right)},\\   
       \beta_\pm&= \Lambda  R^2\frac{(3a_0 \pm 1)}{3 (a_0\pm1)},\\
           \beta_\pm&=\frac{c_1}{2 (1\mp x_0)},
\end{align}
respectively.
The relation for $\beta_\pm$ and the parameters $x_0$, $\Omega^2$, $c_1$ and $a_0$ maybe inverted and used to express the functions $P^2$ in terms of $\beta_\pm$.
In turn, the frame coefficients read
\begin{align}
    P^2&=-\frac{\left(x^2-1\right) (12 \beta_- \beta_++\beta_- \Lambda  R^2
   (x-1) (x+5)+\Lambda  R^2 (x+1) (\beta_+ (x-5)-2 \Lambda  R^2 (x-1)))}{3 \beta_- (x+1)^2+3 \beta_+ (x-1)^2-2 \Lambda  R^2 \left(3 x^2+1\right)},\\
    P^2&= \frac{\left(x^2-1\right) \left(\sqrt{\beta_-^2-\beta_-
   \beta_++\beta_+^2}-\beta_--\beta_+\right) \left(2
   \sqrt{\beta_-^2-\beta_- \beta_++\beta_+^2}+\beta_-
   x+\beta_--\beta_+ x+\beta_+\right)}{2 \sqrt{\beta_-^2-\beta_-
   \beta_++\beta_+^2}+3 \beta_- x+\beta_--3 \beta_+ x+\beta_+},\\
    P^2&=-\frac{4 \beta_- \beta_+ \left(x^2-1\right)}{\beta_- (x+1)^2+\beta_+
   (x-1)^2}.
\end{align}
Finally, the angle deficits $\delta_\pm$, as defined in \cite{PodolskyMatejov2021,PodolskyMatejov2022} are
\begin{equation}
    \delta_\pm=2\pi(1+\mathcal{C}_\pm).
\end{equation}

By expressing $\beta_\pm$ in terms of $\delta_\pm$, we obtain full agreement with our solutions and the Podolský-Matejov results in the non-charged limit, as expected.
In particular our generic solution with $\Lambda\neq0$ recovers the generic solution given by $(19)$ in \cite{PodolskyMatejov2022}, while our special $\Lambda=0$ recovers the $\Lambda=0$ case given by $(23)$ in \cite{PodolskyMatejov2022} and $(65)$ in \cite{PodolskyMatejov2021}.

Incidentally the above procedure is convenient for recovering the limit when no conical singularities are present.
Solutions to EEH on a smooth two-sphere without the singularities were already studied in \cite{Buk}.
The non-singular limit is obtained by setting $\beta_+=\beta_-=1$.
The rotating $\Lambda\neq0$ solution \eqref{generalP} corresponds to the generic solution from \cite{Buk} given by $(38)$ therein, while
the rotating $\Lambda=0$ solution \eqref{Pextr0} corresponds to the $\Lambda=0$ limit of $(38)$.
Finally, both non-rotating $\Lambda\neq0$ solutions correspond to the non-rotating solution given by $(54)$ in \cite{Buk}.

\section{Summary}

In this work, we provide a novel construction of extremal isolated horizons that admit a Hopf bundle (or a higher Chern class bundle) structure, under the assumption that the null generators are transverse to the fibers. Our analysis extends the previous result obtained by Dobkowski-Ry\l{}ko, Lewandowski, and Ossowski \cite{DLO} where the type D equation was solved for isolated horizons of such structure. Here, instead, we address the Einstein vacuum extremal horizon equation with cosmological constant (\ref{extremity2}) and find its solutions.

We begin with the construction of an extremal isolated horizon of a $U(1)$-principal fiber bundle over $S^2$. 
On the axisymmetric data induced on the space of null generators -- consisting of the 2-metric $g$ and 1-form $\omega$ -- we impose the EEH equation with a cosmological constant and determine its solutions.
These are classified into three distinct families (Table \ref{table1}): rotating $\Lambda\neq0$ solution, rotating $\Lambda=0$ solution, and non-rotating solution. We show that no non-rotating solutions exist when the cosmological constant vanishes. 
For each family, we provide the admissible ranges of parameters; in particular, for the general case, these are illustrated in Fig.~\ref{fig:parameters}. 
The generic solution, apart from the cosmological constant $\Lambda$, is characterized by three parameters: area-radius parameter $R^2$, together with $\Omega$ and $x_0$.

In the second part of the paper, we focus on the embeddability of the found solutions in the Pleba\'{n}ski-Demia\'{n}ski, which admit horizons of the structure studied in this work. We provided embeddings of both the rotating and non-rotating solutions, identifying our parameters with horizon radius $r_H$, Kerr and NUT parameters $a$ and $l$, respectively, and acceleration $\alpha$, except for the recently obtained accelerating Taub-NUT-de Sitter spacetime. Furthermore, our extremal solutions satisfy the Petrov type D equation; however, generically, they will not be of the type D. By comparing the frame coefficients for both 2-metrics, we establish explicit relations between the parameters of those solutions.

We also connect our solution with several known families of extremal horizons. In particular, the generic axially symmetric solution found by Lucietti-Kunduri \cite{Lucietti2013}.We verify that our general solution~(\ref{generalP}) coincides with theirs, as expected. Therefore, the intrinsic geometry $(g, \omega)$ found by Lucietti-Kunduri, restricted by imposing conditions on the frame coefficient $P^2$ \eqref{smoothness} can be used to construct a smooth extremal isolated horizon of the non-trivial topology considered in this work. More recently, Podolský and Matejov \cite{PodolskyMatejov2021, PodolskyMatejov2022} classified extremal isolated horizons with conical singularities using the Newman-Penrose formalism and studied their embeddability in type D spacetimes. 
We show that our generic solutions, with both vanishing and non-vanishing cosmological constant, coincide with the solutions given in \cite{PodolskyMatejov2021} and \cite{PodolskyMatejov2022}, respectively. Finally, our solutions have a proper limit to the smooth solutions of Buk and Lewandowski \cite{Buk} to the EEH equation with cosmological constant.
 
\vspace{3mm}


\noindent\textbf{Acknowledgements:} We would like to thank Adam Szereszewski and Wojciech Kami\'nski for insightful discussions. This research was supported by the Polish National Science Centre grant No. 2021/43/B/ST2/02950.


%

\end{document}